\documentclass[aps,prl,twocolumn,showpacs,superscriptaddress,groupedaddress]{revtex4-1}  
\usepackage{graphicx}  
\usepackage{dcolumn}   
\usepackage{bm}        
\usepackage{amssymb}   
\usepackage{amsmath}   
\usepackage{braket}    
\usepackage{hyperref}
\usepackage{natbib}

\begin{document}

\title{Proposal for a quantum random number generator using coherent light and a non-classical observable}
\author{Christopher C. Gerry$^{1}$, Richard J. Birrittella$^{2}$, Paul M. Alsing$^{2}$, Amr Hossameldin$^{3}$\\
	 Miller Eaton$^{3}$ and Olivier Pfister$^{3}$\\
\textit{$^{1}$Department of Physics and Astronomy, Lehman College,\\
The City University of New York, Bronx, New York, 10468-1589,USA} \\
\textit{$^{2}$Air Force Research Laboratory, Information Directorate, Rome, NY, USA, 13441}\\
\textit{$^{3}$Department of Physics, University of Virginia, Charlottesville, VA 22904, USA}
}

\date{\today}

\begin{abstract}
The prototype quantum random number (random bit) generators (QRNG) consists of one photon at a time falling on a $50:50$ beam splitter followed by random detection in one or the other other output beams due to the irreducible probabilistic nature of quantum mechanics.  Due to the difficulties in producing single photons on demand, in practice, pulses of weak coherent (laser) light are used.  In this paper we take a different approach, one that uses moderate coherent light.  It is shown that a QRNG can be implemented by performing photon-number parity measurements.  For moderate coherent light, the probabilities for obtaining even or odd parity in photon counts are $0.5$ each.  Photon counting with single-photon resolution can be performed through use of a cascade of beam splitters and single-photon detectors as was done recently in a photon-number parity-based interferometry experiment involving coherent light. We highlight the point that unlike most quantum-based random number generators, our proposal does not require the use of classical de-biasing algorithms or post-processing of the generated bit sequence.
\end{abstract}

\maketitle

\section{\label{sec:Intro} Introduction}

\noindent As is well known, there is a need for generating random numbers (random bits) in many areas of science and engineering such as for Monte-Carlo simulations and secret communications.  In the latter application, a sequence of random bits can be used to establish a key.  However, most random number generators are based on mathematical algorithms that are entirely deterministic and thus produce only a sequence of pseudo-random numbers.  On the other hand, quantum mechanics offers the prospect of using a physical, nondeterministic, quantum process that can produce a truly random bit sequence.  The quantum world, as described by standard quantum mechanics, is irreducibly probabilistic.  \\

\noindent Recently, Herrero-Collantes and Garcia-Escarttin \cite{ref:Herraro} have reviewed a wide range of quantum random number generators, many of which are quantum-optical \cite{ref:QRNGbook}. Optical schemes for the realization of a quantum random number generator (QRNG) generally involve non-classical states of light.  In this paper, we propose a scheme that employs quasi-classical light (light prepared in a coherent state) but where an observable with no classical analog is used; that observable being the photon-number parity operator.  \\

\noindent The prototype of the optical QRNG is, as is well known \cite{ref:Jennewien,ref:Stefanov,ref:Wang}, the action of a lossless $50:50$ beam splitter upon a single incident photon.  With the input state $\ket{\psi_{\text{in}}}=\ket{1}_{a}\otimes\ket{0}_{b}$ the output state is $\ket{\psi_{\text{out}}} = \tfrac{1}{\sqrt{2}}\left(\ket{1}_{a'}\otimes\ket{0}_{b'} + e^{i\delta}\ket{0}_{a'}\otimes\ket{1}_{b'}\right)$ where the labels $a,b$ denote the input modes while $a',b'$ the output modes.  The first term represents a transmitted photon and the second a reflected one.  The phase $\delta$ depends on the construction of the beam splitter.  For example in the typical case where the reflected beam picks up a $\pi/2$-phase shift, we have $\delta=\pi/2$.  However, the specific value of the phase-shift is irrelevant for the purpose of quantum random number generation as no interference effects are involved.  The detection scheme probes the particle nature of light.  Single photon detectors are placed at the outputs of the beam splitter perform state reductive measurements to project out the transmitted photon state $\ket{1}_{a'}\otimes\ket{0}_{b'}$ fifty percent of the time or the reflective photon state $\ket{0}_{a'}\otimes\ket{1}_{b'}$ fifty percent of the time.  The path taken by any given photon prior to detection is objectively indefinite, this being a manifestation of the irreducible probabilistic nature of the quantum world.  The location of the photon revealed by the clicking of one of the detectors constitutes a projective measurement on $\ket{\psi_{\text{out}}}$ onto one or the other outcomes, though which detector actually fires is a random event.  Therefore a sequence of single photon detection events will have random outcomes; this is the basis for building a QRNG.  The single-photon $50:50$ beam splitter device would make for an ideal QRNG but for the fact that reliable and efficient on-demand single-photon sources are not yet available.  Implemented versions of this approach instead use attenuated laser pulses \cite{ref:Jennewien,ref:Stefanov,ref:Wang} and hence really only provide classical random number generation that mimics, to some extent, a true single-photon QRNG. An attenuated laser beam, as described by a weak amplitude coherent state

\begin{equation}
	\ket{\alpha} = e^{-|\alpha|^{2}/2}\sum_{n=0}^{\infty}\frac{\alpha^{n}}{\sqrt{n!}}\ket{n}
	\simeq \ket{0} + \alpha\ket{1} + ....\;\;\;\;\;\; |\alpha| \ll 1,
	\label{eqn:intro_1}
\end{equation}

\noindent so that the input state of the beam splitter is $\ket{\alpha}_{a}\otimes\ket{0}_{b}\simeq \ket{0}_{a}\otimes\ket{0}_{b} + \alpha\ket{1}_{a}\otimes\ket{0}_{b}$.  This is dominated by the double vacuum state $\ket{0}_{a}\ket{0}_{b}$ as the probability for finding one photon goes as $|\alpha|^{2} \ll 1$ so that most of the time there are no photons present at all.  Single-photon states are not realizable by pulses of weak coherent light because of the dominance of the vacuum state in the weak field limit.  Recall that the coherent state $\ket{\alpha}$ is a classical-like state for all values of the parameter $\alpha$ in the sense that the corresponding Glauber-Sudarshan $P$-function \cite{ref:Glauber,ref:Sudarshan} is a delta function \cite{ref:Hillery}.  In contrast, the one photon Fock state $\ket{1}$ is a highly non-classical state of whose $P$-function is a second order derivative of a delta function \cite{ref:GerryBook}.  Moreover, the Wigner function of the one photon state becomes negative in some regions of phase space.  Issues in developing "on demand" single-photon sources have been reviewed by Scheel \cite{ref:Scheel}.  \\

\noindent One way to implement a one-photon QRNG is to use a non-degenerate down-converter whose output is weak enough so that at most only the double-vacuum and the biphoton states are populated, i.e. a state of the form $\gamma\ket{0}_{s}\otimes\ket{0}_{i} + \eta\ket{1}_{s}\otimes\ket{1}_{i}$ where $|\gamma|^{2} \gg |\eta|^{2}$ and where the labels $s,i$ denote the signal and idler modes, respectively.  One could then use the detection of the idler photon to herald the signal photon falling on the $50:50$ beam splitter whose subsequent detection determines the value of the bit, as described above.  The downside of this approach is that the most probable output state of the down-converter is the twin-vacuum state $\ket{0}_{s}\otimes\ket{0}_{i}$.  Furthermore, the conversion of pump photons into pairs of single-photon states is low: typically on the order of $1/10^{8}$.  Thus the production rate of photon pairs is limited, hence limiting the rate of production of strings of random bits.  By increasing the amplitude of the pump field, one runs the risk of generating the state $\ket{2}_{s}\otimes\ket{2}_{i}$ or even higher number twin-Fock states, which spoils the usefulness of the setup as a QRNG.\\

\noindent Kwon \textit{et al.} \cite{ref:Kwon} have examined a variation on this setup that uses the Hong-Ou-Mandel (HOM) effect \cite{ref:Hong}.  The randomness in this case comes from the randomness of the output paths taken together by two photons after each simultaneously falls on opposite sides of a $50:50$ beam splitter.  The outputs of this beam splitter are allowed to fall on secondary $50:50$ beam splitters whose outputs will result in two photons detected in one beam or the other as per the HOM effect.  One is required to detect one photon on each of the output beams of the secondary beam splitters in order to guarantee that the two photons were either transmitted or reflected together at the first beam splitter.  However, the issue of the production of twin single-photons states with a weak enough pump field to exclude the production of the two photon twin Fock states is the same as in the previous discussion. \\

\noindent The states of light described above are Fock (or number) states of single-photon occupation.  These states are highly non-classical states of light.  Note that these single-photon states are, in terms of their generation, secondary in that they are produced by spontaneous down-conversion through pumping by a light field prepared in a moderate coherent state (laser light) usually approximated as a classically prescribed light field.  This begs the question: Is it possible to directly use a strong, or even "medium", strength laser light field, assumed to be prepared in a coherent state, to implement a QRNG? In the balance of this paper we answer this question in the affirmative.  \\

\section{\label{sec:Parity_QRNG} Quantum Random Number Generation based on Photon-Number parity measurements}

\noindent Our approach requires only moderate quasi-classical light field, i.e. a field prepared in a coherent state by a phase-stabilized laser \cite{ref:Cahill} (though even this condition will be later relaxed), and does not depend on beam splitting and on the path of the light propagation.  Instead, it relies on the measurement of a quantum mechanical observable having no classical analog: photon-number parity \cite{ref:Birrittella}, which is simply the evenness or oddness of photon numbers.  The scheme is based on the fact that for a sufficiently large enough coherent state amplitude the average parity for the coherent state is zero, leading to the probability of producing the outcome ``even" or "odd" under photon-number parity measurements being equalized to $P_{\text{even}} = P_{\text{odd}}=1/2$.\\

\noindent A single-mode quantized field is described by a set of annihilation and creation operators $\hat{a}$ and $\hat{a}^{\dagger}$ satisfying the commutation relation $\left[\hat{a},\hat{a}^{\dagger}\right]=1$ with photon-number operator $\hat{n}_{a}=\hat{a}^{\dagger}\hat{a}$ satisfying $\hat{n}_{a}\ket{n}_{a}=n\ket{n}_{a}$ where $n \in \mathbb{Z}^{\;0+}$, understood as the number of photons associated with the Fock state $\ket{n}$.  The canonical coherent state for the field is given by 

\begin{equation}
	\ket{\alpha} =e^{-\tfrac{1}{2}|\alpha|^{2}}\sum_{n=0}^{\infty}\frac{\alpha^{n}}{\sqrt{n!}}\ket{n},
	\label{eqn:1}
\end{equation}

\noindent and may be defined as either a right eigenstate of the annihilation operator, $\hat{a}\ket{\alpha}=\alpha\ket{\alpha}$, or as the displaced vacuum state, $\ket{\alpha}=\hat{D}\left(\alpha\right)\ket{0}$. where $\hat{D}\left(\alpha\right)=e^{\alpha\hat{a}^{\dagger}-\alpha^{*}\hat{a}}$ is the usual displacement operator.  The average photon number $\bar{n}$ for the coherent state is found to be $\bar{n}_{\text{coh}}=\braket{\alpha|\hat{n}|\alpha}=|\alpha|^{2}$.  The photon number distribution for the coherent state is the Poissonian, i.e. $P_{m}=|\braket{m|\alpha}|^{2}=e^{-\bar{n}}\left(\bar{n}^{m}/m!\right)$ with, by definition, a photon-number variance $\Delta^{2}\hat{n} = \braket{\hat{n}^{2}} - \braket{\hat{n}}^{2}=\bar{n}$ where the moments of the number operator are given by 

\begin{equation}
	\braket{\hat{n}^{k}} = \sum_{n=0}^{\infty}n^{k}P_{n}, \;\;\;\;\;\;k=1,2..\;.
	\label{eqn:2}
\end{equation} 

\noindent The photon number parity operator of a single-mode field, denoted $\hat{\Pi}$, can be defined in a number of equivalent ways:

\begin{align}
	\hat{\Pi} &= \left(-1\right)^{\hat{n}} = e^{i\pi\hat{n}} = \hat{P}_{e} - \hat{P}_{o} \nonumber \\
	&= \sum_{n=0}^{\infty}\cos\left(n\pi\right)\ket{n}\bra{n},
\end{align}

\noindent where $\hat{P}_{e\left(o\right)}$ are the even (odd) parity projection operators given by 

\begin{align}
	\hat{P}_{e} &= \sum_{n=0}^{\infty} \cos^{2}\left(n\pi/2\right)\ket{n}\bra{n} = \sum_{m=0}^{\infty}\ket{2m}\bra{2m}  ,\nonumber \\
	& \label{eqn:3} \\
	\hat{P}_{o} &= \sum_{n=0}^{\infty} \sin^{2}\left(n\pi/2\right)\ket{n}\bra{n} = \sum_{m=0}^{\infty}\ket{2m+1}\bra{2m+1}  \nonumber,
\end{align}

\noindent where $\hat{P}_{e}+\hat{P}_{o}\equiv\hat{I}$ as per the positive operator-valued measure (POVM) unity condition.  Note that the expectation value of the parity operator contains within it all moments of the number operator:

\begin{equation}
	\braket{\hat{\Pi}} = \braket{e^{i\pi\hat{n}}} = \sum_{n=0}^{\infty}\frac{\left(i\pi\right)^{n}}{n!}\braket{\hat{n}^{n}}.
	\label{eqn:4}
\end{equation}

\noindent This operator has been discussed in many contexts in quantum optics; for example, in connection with proposed tests of Bell-type inequalities \cite{ref:Banaszek} and in connection with quantum optical interferometry \cite{ref:Mimih}.  Furthermore, the Wigner function quasi-probability distribution $W\left(\beta\right)$ may be given in terms of the expectation value of the displaced parity operator \cite{ref:Cahill}

\begin{equation}
	W\left(\beta\right) = \tfrac{2}{\pi}\braket{\hat{D}\left(\beta\right)\hat{\Pi}\hat{D}^{\dagger}\left(\beta\right)}.
	\label{eqn:5}
\end{equation}

\noindent The expectation value of the parity operator itself is, up to a proportionality constant, the value of the Wigner function at the origin of phase space \cite{ref:Banaszek2}

\begin{equation}
	W\left(0\right) = \frac{2}{\pi}\sum_{n=0}^{\infty}\left(-1\right)^{n}p_{n}\;\;\;\to\;\;\;\braket{\hat{\Pi}} = \frac{\pi}{2}W\left(0\right),
	\label{eqn:added}
\end{equation}

\noindent where $p_n$ are the $n$-photon probabilities.  For a coherent state $\ket{\alpha}$, the expectation value of the parity operator is 

\begin{equation}
	\braket{\alpha|\hat{\Pi}|\alpha} = \braket{\hat{\Pi}}_{\alpha} = e^{-\bar{n}}\sum_{n=0}^{\infty}\frac{\left(-\bar{n}\right)^{n}}{n!} = e^{-2\bar{n}}.
	\label{eqn:6}
\end{equation}

\noindent On the other hand, the expectation values of the even and odd projection operators with respect to a coherent state yield the probabilities 

\begin{align}
	P_{e} = \braket{\hat{P}_{e}}_{\alpha} &=  e^{-\bar{n}}\sum_{m=0}^{\infty}\frac{\bar{n}^{2m}}{\left(2m\right)!} \nonumber \\
	&= e^{-\bar{n}}\cosh\bar{n} = \tfrac{1}{2}\left(1+e^{-2\bar{n}}\right),  \label{eqn:7}\\
	&  \nonumber\\
	P_{o} = \braket{\hat{P}_{o}}_{\alpha} &=  e^{-\bar{n}}\sum_{m=0}^{\infty}\frac{\bar{n}^{2m+1}}{\left(2m+1\right)!} \nonumber \\
	&= e^{-\bar{n}}\sinh\bar{n} = \tfrac{1}{2}\left(1-e^{-2\bar{n}}\right) \label{eqn:8} . 
\end{align}

\noindent In the limit of large $\bar{n}$ one has $\braket{\hat{\Pi}}_{\alpha}\to 0$ as seen from Eq.~\ref{eqn:6} and the probabilities $P_{e},\;P_{o} \to 1/2$ rapidly as $\bar{n}$ becomes large.  In fact, the average photon number need not be very large at all.  For example, for $\bar{n} = 6$, the probabilities become $P_{e}=P_{o}=1/2$ rounded to five decimal places. For $\bar{n}=16$, this is true when rounding to thirteen decimal places.  Thus for even moderately large $\bar{n}$ the probabilities for detecting even or odd photon numbers are highly balanced.  In other words, for large enough $\bar{n}$, there will be no bias in the outcomes of parity measurements.  In contrast, generating random digits from single photons on a beam splitter depends on the ability to manufacture a perfect $50:50$ beam splitter.  \\

\noindent In the preceding we assumed the availability of a pure coherent state as provided by a phase-stabilized laser.  But a statistical mixture of coherent states of identical amplitudes would suffice.  Consider the phase-averaged coherent state \cite{ref:Allevi} 

\begin{equation}
	\rho_{\text{coh}} = \frac{1}{2\pi}\int_{0}^{2\pi}d\varphi\ket{re^{i\varphi}}\bra{re^{i\varphi}},
	\label{eqn:9}
\end{equation}

\noindent where $r=|\alpha|=\sqrt{\bar{n}}$.  It is easily shown 

\begin{align}
	\braket{\hat{\Pi}} = \text{Tr}\left[\rho_{\text{coh}}\hat{\Pi}\right] &= \frac{1}{2\pi}\int_{0}^{2\pi}d\varphi\sum_{n=0}^{\infty}\left(-1\right)^{n}|\braket{n|re^{i\varphi}}|^{2} \nonumber \\
	&= e^{-\bar{n}}\sum_{n=0}^{\infty}\frac{\left(-\bar{n}\right)^{n}}{n!} = e^{-2\bar{n}},
	\label{eqn:10}
\end{align}

\noindent which is identical to Eqn.~\ref{eqn:6}.  \\

\noindent This process will not work for any form of light.  For example, thermal light is not a viable alternative to the use of coherent states (or mixtures of coherent states) for a parity-measurement based QRNG.  Thermal light is described by the well known density operator \cite{ref:GerryBook}

\begin{equation}
	\rho_{\text{therm}} = \frac{1}{1+\bar{n}}\sum_{n=0}^{\infty}\left(\frac{\bar{n}}{1+\bar{n}}\right)^{n}\ket{n}\bra{n},
	\label{eqn:11}
\end{equation}

\noindent where $\bar{n} = \left(e^{\hbar\omega/k_{B}T}-1\right)^{-1}$.  The photon number distribution in this case is 

\begin{equation}
	P_{m} = \braket{m|\rho_{\text{therm}}|m} = \frac{1}{1+\bar{n}}\left(\frac{\bar{n}}{1+\bar{n}}\right)^{m}, 
	\label{eqn:12}
\end{equation}

\noindent and thus the average parity for this state is given by 

\begin{equation}
	\braket{\hat{\Pi}}_{\text{thermal}} =  \frac{1}{1+\bar{n}}\sum_{n=0}^{\infty}\left(-\frac{\bar{n}}{1+\bar{n}}\right)^{n} = \frac{1}{1+2\bar{n}},
	\label{eqn:13}
\end{equation}

\noindent which goes to zero only asymptotically in the limit that $\bar{n}\to \infty$.  This is because the photon number distribution for thermal light is super-Poissonian, i.e. broader than the corresponding Poissonian distribution associated with light in a coherent state with the same average photon number.  In the latter case, the photon number fluctuations are given by $\Delta^{2}\hat{n} = \bar{n}$ whereas for super-Poissonian light $\Delta^{2}\hat{n} > \bar{n}$.  For thermal light in particular, $\Delta^{2}\hat{n} = \bar{n}^{2} + \bar{n}$ \cite{ref:GerryBook}.  Therefore, with regards to easily generated classical-like states of light, only pure coherent states of light or phase-averaged coherent states of the form Eqn.~\ref{eqn:9}, where the average photon number need not be very high, are usable for a QRNG based on photon-number parity measurements.  

\section{\label{sec:exp}Experimental Considerations and Implementation}

\noindent Here we seek to address some experimental considerations in utilizing the outlined proposal. As the proposed quantum random number generator requires the capacity to distinguish between different photon numbers, a reliable and efficient photon-number resolving (PNR) detector is required.  Recent work done by Maga\~na-Loaiza \textit{et al.} \cite{ref:Magana} has shown photon-number counting has been done up to six photons using transition edge sensors (TES).  The authors demonstrated quantum-state engineering of an entangled two-mode state of light with up to ten photons exhibiting nearly Poissonian photon statistics. Furthermore, recent work has demonstrated the potential for a TES to resolve up to 16 photons \cite{ref:Morais}.  A description of the TES detector involved can be found in \cite{ref:Lita2,ref:Gerrits,ref:Schmidt}.  One drawback of such detectors, however, is that they suffer from long rest times of $\sim10\mu$s \cite{ref:Lita2}.  Conversely, silicon nanowire detectors, which can be used for PNR detection \cite{ref:Achilles,ref:Jahanmirinejad} either directly \cite{ref:Cahall} or through multiplexing \cite{ref:Tao} are significantly faster and can reach detection rates above 100 MHz. Moreover depending on the measurement implementation, photon-number measurements can substantially exceed kHz range. Ref \cite{ref:Cohen2} can exceed MHz range with deadtimes on the order of 50 ns, which is consistent with other superconducting nanowire single photon detector (SNSPD) multiplexing \cite{ref:Cahall,ref:Endo}.\\

\noindent Further, the long detector rest time of TES can be somewhat circumvented through the use of modulo 4 measurement binning, provided one has high photon number resolution and a coherent state of high enough average photon number.  Because there are four possible outcomes, this method can produce two random bits for every measurement, thus doubling the bit generation rate by assigning a bit string 00,01,10,11 to the respective potential outcomes.  Here we can define new POVM elements $\hat{P}^{\left(k\right)}$, analogous to Eq.~\ref{eqn:3}, as 

\begin{equation}
	\hat{P}^{\left(k\right)} = \sum_{m=0}^{\infty}\ket{4m+k}\bra{4m+k},
	\label{eqn:mod4_0a}
\end{equation}

\noindent such that $\sum_{k}\hat{P}^{\left(k\right)}\equiv\hat{I}$ and the even/odd projectors are given by $\hat{P}_{e} = \hat{P}^{\left(k=0\right)}+\hat{P}^{\left(k=2\right)}$ and $\hat{P}_{o} = \hat{P}^{\left(k=1\right)}+\hat{P}^{\left(k=3\right)}$, respectively. Likewise, we now have 

\begin{align}
	\hat{\Pi} &= \hat{P}_e - \hat{P}_o \nonumber \\
	&=\hat{P}^{\left(0\right)}+\hat{P}^{\left(2\right)} - \left(\hat{P}^{\left(1\right)}+\hat{P}^{\left(3\right)}\right).
	\label{eqn:mod4_0b}
\end{align}
\noindent These new POVM elements can be thought of as corresponding to a form of higher order parity where, for example, $\hat{P}^{\left(k=0\right)}$ is an even-state projection of every \textit{other} even photon number and likewise for $\hat{P}^{\left(k=1\right)}$ for the odd-state projections.  If one performs such a measurement, the outcome will have remainders of 0,1,2, or 3.  The probability to measure a remainder $k$ is given by  

\begin{align}
	P_{\text{mod}_4}^{\left(k\right)} = \braket{\hat{P}^{\left(k\right)}} &= e^{-\bar{n}}\sum_{n=0}^{\infty}\frac{\bar{n}^{4n+k}}{\left(4n+k\right)!} \nonumber \\
	&= \frac{1}{4}\left(1+2e^{-\bar{n}}\cos\left(\bar{n}-\tfrac{k\pi}{2}\right)+\left(-1\right)^{k}e^{-2\bar{n}}\right),
	\label{eqn:mod4_1}
\end{align}

\noindent which approaches a uniform distribution as $e^{-\bar{n}}\to 0$.  If one has the ability to measure larger photon numbers, this this will double generation rates of unbiased random numbers.  For example, a coherent state of mean photon number $\bar{n}=10$ results in a bias on the order of $10^{-5}$ and a mean photon number of $\bar{n}=16$ results in a bias of $10^{-8}$.  \\

\noindent We also point out that our proposal is remarkably robust to experimental imperfections including photon loss, phase and amplitude fluctuations and detector inefficiency.  Detector efficiency is typically modeled via beamsplitter transformation with a efficiency coefficient $\eta$ modeling the beamsplitter reflectivity as $\sqrt{1-\eta}$, where a partial trace is then performed over the reflected mode.  Due to the nature of coherent states, this yields a coherent state of weaker amplitude.  For a beamsplitter $\hat{B}=e^{\theta\left(\hat{a}^{\dagger}\hat{b} - \hat{a}\hat{b}^{\dagger}\right)}$, with $\sqrt{\eta}=\cos\theta$ we have

\begin{equation}
	\text{Tr}_b\left[\hat{B}\ket{\alpha}_a\bra{\alpha}\otimes\ket{0}_b\bra{0}\hat{B}^{\dagger}\right] = \ket{\sqrt{\eta}\alpha}_a.
	\label{eqn:robust_1}
\end{equation}

\noindent Provided one calibrates the overall loss in the experiment, it is possible to increase the starting value of $|\alpha|$ to compensate for any loss and detector inefficiency. \\

\noindent We can also investigate the impact of phase and amplitude noise.  For a density matrix

\begin{equation}
	\rho = \int_{0}^{2\pi}d\phi\ket{re^{i\phi}}\bra{re^{i\phi}}f\left(\phi\right),
	\label{eqn:robust_2}
\end{equation}

\noindent where $f\left(\phi\right)$ is a general normalized function and $r=\sqrt{\bar{n}}$. We evaluate the expectation value of the parity operator as 

\begin{align}
	\braket{\hat{\Pi}} &= \text{Tr}\left[\rho\;e^{i\pi\hat{n}}\right] \nonumber \\
	&= \int_{0}^{2\pi}d\phi\;f\left(\phi\right)\sum_{n=0}^{\infty}\braket{n|re^{i\phi}}\braket{re^{i\phi}|e^{i\pi\hat{n}}|n}\nonumber \\
	&= \int_{0}^{2\pi}d\phi\;f\left(\phi\right)\sum_{n=0}^{\infty}\left(-1\right)^{n}|\braket{n|re^{i\phi}}|^2 \nonumber \\
	&= \int_{0}^{2\pi}d\phi\;f\left(\phi\right)e^{-\bar{n}}\sum_{n=0}^{\infty}\frac{\left(-\bar{n}\right)^n}{n!} \nonumber \\
	&= e^{-2\bar{n}}\int_{0}^{2\pi}d\phi\;f\left(\phi\right)\nonumber \\
	&= e^{-2\bar{n}},
	\label{eqn:robust_3}
\end{align}

\noindent which is exactly the same as one would get without phase noise.  Now instead suppose we have 

\begin{equation}
	\rho = \int_{\sqrt{\bar{n}-\Delta}}^{\sqrt{\bar{n}+\Delta}}d\alpha\ket{\alpha}\bra{\alpha}f\left(\alpha\right),
	\label{eqn:robust_4}
\end{equation}

\noindent one would find for the expectation value of the parity operator

\begin{equation}
	\braket{\hat{\Pi}} = \text{Tr}\left[\rho\;e^{i\pi\hat{n}}\right] = \int_{\sqrt{\bar{n}-\Delta}}^{\sqrt{\bar{n}+\Delta}}d\alpha f\left(\alpha\right)e^{-2|\alpha|^2}.
	\label{eqn:robust_5}
\end{equation}

\noindent If we assume that $f\left(\alpha\right)$ is, at worst, a delta function centered at $\alpha=\sqrt{\bar{n}-\Delta}$, we can directly evaluate the integral above to get $\braket{\hat{\Pi}}=e^{-2\left(\bar{n}-\Delta\right)}$, which goes to zero for large enough $\bar{n}$ and small enough $\Delta$.  Otherwise, $f\left(\alpha\right)$ will put more weight on factors with larger $\alpha$ which would result in a smaller $\braket{\hat{\Pi}}$. \\

\noindent 
While we have not conducted a rigorous security analysis against adversarial attacks,
our proposal is secure in the sense that an eavesdropper siphoning photons from our laser source will generate (potentially) random numbers uncorrelated to what is generated by the experimenter.  Because of the nature of coherent states, while tampering may go unnoticed, it does not effect the randomness of the produced bit sequence, provided the coherent state intensity is such that $\braket{\hat{\Pi}}=0$ remains true. As another advantage to our proposal, current random number generators with high generation rates require classical de-biasing algorithms (post-processing of the data), which by virtue of being classical is not based in true randomness (i.e. pseudo-random).  This is due to the fact that imperfections in current QRNG approaches, such as detector inefficiency and non-deterministic single-photon sources, lead to quantum randomness following non-uniform distributions that must be corrected with classical de-biasing \cite{ref:Herraro}.  
In principle, our proposal could be implemented with a laser and a single detector, since it only involves counting the total number of photons impinging on the detector. In addition, our proposal is robust to photon loss, since loss corresponds to the mapping of coherent states to smaller amplitude coherent states. In other QRNG implementations using beamsplitters and multiple detectors, a potential issue may arise due to the backscattering of photons from one detector branch to another. This could produce undesirable correlations which have to be removed, for example, through post-processing. However, in our proposal, only the total number of photons detected determines the random bit sequence, not \textit{which} detectors fire, even in the presence of backscatter.
Thus, our proposal has the potential to be more secure.   \\

\section{\label{sec:Parity_Detect}Further Remarks on Photon-Number Parity Detection}

\noindent To complete the description of the proposed QRNG there needs to be a discussion on the measurement of photon-number parity.  Ideally one would want to perform a kind of measurement that would reveal photon-number parity directly, that is, without first determining the photon number.  In principle, this could be done by a quantum non-demolition (QND) measurement of the type discussed by Gerry \textit{et al.} \cite{ref:Gerry2}.  The proposed measurement scheme of that paper amounts to an extension of the proposed QND scheme for the measurement of photon number \cite{ref:Munro}.  But both of these proposals require large cross-Kerr interaction which have yet to be made available in the laboratory despite attempts to engineer the required large third-order nonlinear susceptibilities $\chi^{\left(3\right)}$ by the techniques of electromagnetically induced transparency \cite{ref:Fleischhauer}. \\

\noindent The obvious alternative is to perform photon number counting at a resolution of one photon and calculate the parity of each event by raising $-1$ to the number of measured photons.  Fortunately, such a procedure has already been performed in the laboratory in the context of quantum optical interferometry.  Some years ago, the author \cite{ref:Gerry2} proposed the use of photon-number parity measurements in quantum optical interferometry as an alternative to the usual method of subtracting the two output photocurrents on an interferometer.  This was motivated by the fact that the usual method was insensitive to the induced phase-shift for certain kinds of states such as $N00N$ states, which had been proposed by Dowling and collaborators \cite{ref:Dowling} as a resource for obtaining Heisenberg-limited sensitivites in phase-shift measurements but without a well defined scheme for detecting the phase shift.  In Ref. \cite{ref:Gerry2} it was shown that using photon-number parity measurements with a $N00N$ state of $N$ photons one could obtain Heisenberg-limited sensitivity $\Delta\phi_{\text{HL}} = 1/N$, an improvement in noise reduction over the standard quantum limit $\Delta\phi_{\text{SQL}} = 1/\sqrt{N}$ by a factor of $1/\sqrt{N}$.  The idea of using parity-like measurements in the context of quantum metrology first appeared in a paper by Bollinger \textit{et al.} \cite{ref:Bollinger} in the context of measurements of atomic transition frequencies with a collection of trapped ions.  In this case, the parity operator consisted of $-1$ raised to the number of ions in the excited state.  \\

\noindent Recently, Cohen \textit{et al.} \cite{ref:Cohen1} have demonstrated the predicted super-resolved phase-shift measurements in the laboratory with coherent (laser) light and photon-number parity measurements.  These photon-number parity measurements were obtained using a photon-number resolving detector (a silicon photo-multiplier, SIPM, \textit{Hamamatsu Photonics,} S10362-11-100U) and an array (cascade) of beamsplitters forming an $N$-port interferometer with a single photon-detector element at the outputs \cite{ref:Reck}.  The input coherent states, of up to 4200 photons on average, were generated by a Ti:Sapphire laser producing 780 nm wavelength light in 150 fs length pulses at a rate of 250 KHz.  The average number of photons per pulse was controlled with calibrated neutral density filters.  In a recent paper, Cohen \textit{et al.} \cite{ref:Cohen2} have discussed calibration of single-photon and multiplexed photon-number-resolving detectors, the latter using a cascade of beam splitters with single-photon detectors as in the experiment discussed in Ref. \cite{ref:Cohen1}.  \\

\noindent Finally, we mention that there exists one more possibility for the measurement of photon-number parity in a single-mode field for cases of Gaussian states, which includes coherent states.  Recall that the Wigner function at the origin of phase space is, up to a multiplicative factor, the expectation value of the parity operator.  Thus one needs a reconstruction of the Wigner function only at the origin of phase space for parity measurements.  The Wigner function over the entire phase space is not required.  Plick \textit{et al.} \cite{ref:Plick2} have shown that the parity of a Gaussian state of light (which includes coherent states of light) can be obtained without photon-number counting using only a balanced homodyne technique and an intensity correlation.  Apparently, this can be done for an arbitrary photon flux.  In Ref. \cite{ref:Cohen2} the authors show that the detection of the parity can be determined independently of the parity operator and that this 'parity-by-proxy' measurement has the same signal as the method of performing a photon-number resolving count and raising $-1$ to that power.  We are not aware of any laboratory implementation of this approach. \\

\section{\label{sec:Parity_Conclusion} Conclusion}
 
\noindent In summary, in this paper we have proposed a quantum random number generator based on the measurement of photon-number parity in the light of a phase-stabilized laser or from laser light prepared in a statistical mixture of coherent states of the same amplitude but with different phases.  

\section{\label{sec:Parity_Acknowledgements} Acknowledgements}

\noindent RJB acknowledges support from the National Research Council Research Associate Program (NRC RAP).
CCG acknowledges support under AFRL Summer Faculty Fellowship  Program (SFFP).
PMA and CCG acknowledge support from the Air Force Office of Scientific Research (AFOSR).  ME, AH, and OP were supported by National Science Foundation grants No. DMR-1839175 and No. PHY-1820882, and by Jefferson Lab LDRD project No. LDRD21-17 under which Jefferson Science Associates, LLC, manages and operates Jefferson Lab.  Any opinions, findings and conclusions  or  recommendations  expressed  in  this  material are those of the author(s)
and do not necessarily reflect the views of the Air Force Research Laboratory (AFRL).

\section{\label{sec:disclosure} Disclosures}

\noindent The authors declare no conflicts of interest.

\nocite{apsrev41Control}
\bibliographystyle{apsrev4-1}
\bibliography{QRNG_Parity_bib}

\end{document}